\documentclass[epj]{svjour}
\usepackage{amsmath}
\usepackage{amssymb}
\usepackage{bm}
\usepackage{graphicx}

\begin{document}

\title{On the deflection of light by topological defects in nematic liquid crystals}
\author{Caio S\'atiro
\thanks{\emph{Present address: Dept. de F\'{\i}sica, Universidade Federal Rural de Pernambuco
R. Dom Manoel de Medeiros s/n, 52171-900, Recife/PE, Brazil}}
\and  Fernando Moraes}
\institute{Departamento de F\'{\i}sica,
           Universidade Federal da Para\'{\i}ba\\
           Caixa Postal 5008, 58051-970,
            Jo\~ao Pessoa, PB, Brazil\\ }

\abstract{
The influence of controlable parameters like temperature and wavelegth on the trajectories of light in a nematic liquid crystal with topological defects is studied through  a geometric model. The model incorporates phenomenological details as how the refractive indices depend on such parameters. The deflection of light by the topological defect is then shown to be greater at lower temperatures and shorter wavelengths.
\PACS{ {61.30.Jf}{Defects in liquid crystals} \and {61.30.Dk}{Continuum models and theories of liquid crystal structure} \and {42.70.Df}{Liquid crystals} \and {78.20.Fm}{Birefringence}}}

\maketitle
\section{Introduction}
Topological defects appear everywhere in physics. Ranging from gravitation and cosmology to Bose-Einstein condensates they are associated to symmetry-breaking phase transitions. In particular, they appear in the isotropic-to-nematic phase transition of calamitic liquid crystals in the form of hedgehogs, disclinations, domain walls or more complicated textures \cite{kleman,repnik}. The deflection of light rays by these defects is an old issue. Grandjean \cite{grand}, already in 1919,  calculated the light paths of extraordinary rays passing by disclination lines \cite{prob}. More recently,  Joets and  Ribotta \cite{joets} used a geometrical model to describe the light propagation in anisotropic and inhomogeneous media like liquid crystals.  Inspired by their work we studied the propagation of light near disclination lines in nematics from a geometric point of view \cite{caio1,caio2,caio3}. In particular, in \cite{caio1}  we observed lensing effects due the deflection of the beams in regions near  the defect. There, we calculated the light trajectories from  a geometry resulting from the application of Fermat's principle\footnote{Kline and Kay \cite{kline} were probably the first to prove that light rays in inhomogeneous anisotropic media are extremals of Fermat's functional.} associated to an effective refractive index $N$ \cite{born} given in terms of $n_o$ and $n_e$, the ordinary and extraordinary refractive index, respectively. In a few words, what we did was to consider the bent light rays as geodesics of a model space of unknown geometry. By identifying Fermat's principle with the geodesic variational principle we were able to find the effective geometries for each defect studied. Knowing the effective geometry, the geodesics were obtained numerically. Incidentally, a geometric model describing elastic properties of nematic liquid crystals which also leads to an effective geometry, appeared recently in the literature \cite{sim}.

Since the refractive indices of a nematic depend both on the temperature and on the wavelenght of the light, a more realistic model incorporating these effects is in order. Li, Gauza and Wu \cite{jun1} modeled the temperature effect on the nematic refractive index based on Vuks equation \cite{vuks} and, by fitting their final expression to experimental data of selected materials, found the unknown coefficients.The same group derived Cauchy formulae (see page 100 of \cite{born}) for the refractive indices of a nematic sample as a function of the wavelenth of the light \cite{jun2}. Again, by fitting their final expression to experimental data, all coeficients were determined. In this work, we incorporate their models to our geometric model for propagation of light in nematics with topological defects, in order to  study temperature and wavelength effects. Although in \cite{caio1} we studied both the symmetric ($k=1$) and asymmetric ($k\neq 1$) defects, without loss of generality, we keep our analysis here mostly for the symmetric cases since their effective geometry is simpler and more intuitive than the asymmetric cases. Since, in all cases, the effective geometry depends only on the ratio $\alpha=n_e/n_o$ it is how the temperature and the wavelength affect this ratio what matters. In section II we review the geometric model and discuss the simplest effective geometry for light traveling by a disclination. In section III we use the results of \cite{jun1} and \cite{jun2} to show how $\alpha$ is affected by temperature and wavelength and study the effect of the variation of this ratio on the light paths near selected defects.

\section{Geometric Model}

Disclinations in nematics are classified according to the topological index (or strength) $k$ which gives a measure of how much the director rotates as one goes around the defect. That is, the director configurations, in the plane $x-y$, are given by \cite{kleman}
\begin{equation}
\varphi(\theta)=k\theta+c , \label{phi}
\end{equation}
where $\varphi$ is the angle  between the molecular axis and the $x$-axis, $\theta$ is the angular polar coordinate and $c=\varphi(0)$. Selected director configurations can be seen on Figure 11.4 of \cite{kleman}. We assume the disclinations are straight and lie along the $z$-axis and the light rays propagate in the $x$-$y$ plane so, effectively, we have a two-dimensional problem.

We consider an optical medium constituted by a  nematic liquid crystal with disclinations \cite{repnik}, where the effective geometry for the light is defined by the line element (equation (25) of \cite{caio1}) 
{\small 
\begin{eqnarray}
& ds^2 & = \left\{n_o^2 \cos^{2}[(k-1)\phi+c]+n_e^2 \sin^{2}[(k-1)\phi+c]\right\}dr^{2} \nonumber\\
& + & \left\{n_o^2 \sin^{2}[(k-1)\phi+c]+n_e^2 \cos^{2}[(k-1)\phi+c]\right\}r^{2}d\phi^{2} \nonumber\\
& - & \left\{2(n_e^2-n_o^2)^{2}\sin[(k-1)\phi+c]\cos[(k-1)\phi+c]\right\}rdrd\phi. \nonumber\\
& & \label{kmetric}
\end{eqnarray}}
The metric (\ref{kmetric}) was obtained by identifying Fermat's principle with the variational principle that determines the geodesics in Riemannian geometry. Let
\begin{equation}
{\cal F}=\int_{A}^{B} N d \ell , \label{fermat}
\end{equation}
where, $d\ell$ is the element of arc length along the path between points $A$ and $B$ and the effective refractive index
\begin{equation}
N^2=n_o^2\cos^2\beta +n_e^2\sin^2\beta , \label{nr}
\end{equation}
where $\beta = (\widehat{\vec{n},\vec{S}})$ is the local angle between the director $\vec{n}$ and the Poynting vector $\vec{S}$. Then, among all possible paths between the generic points $A$ and $B$, Fermat's principle for the extraordinary rays grants us that the path actually followed by the energy is the one that minimizes ${\cal F}$.

In Riemannian geometry the line element $ds$ depends on the position coordinates $x^i$ of the point of the manifold under consideration. That is,
\begin{equation} 
ds^2 = \sum_{i,j}g_{ij}dx^idx^j, \label{riemline}
\end{equation}
where $g_{ij}=g_{ij}(x^i)$ is the metric tensor. The geodesic joinning points $A$ and $B$ in such manifold is obtained by minimizing $\int_A^B ds$, just like Fermat's principle. This leads to a nice interpretation of the light paths as geodesics in an effective geometry \cite{born}.   Thus, we may identify
\begin{equation} 
N^{2}d\ell^2  = \sum_{i,j}g_{ij}dx^idx^j. \label{interp}
\end{equation}
The meaning of this equation is the following: the line element of the optical path, in an Euclidean space with refractive properties, is identified with the line element of an effective geometry characterized by $g_{ij}$.

In \cite{caio2} we showed that the effective geometry for the vortex-like $k=1$, $c=\frac{\pi}{2}$ disclination is that of a cone. The effective metric for this case is obtained by substituting these values in metric (\ref{kmetric}) and rescaling the coordinate $r$ to $\rho=n_e r$. That is, the two-dimensional line element  for this effective geometry, in polar coordinates, is \cite{caio1}
\begin{equation}
ds^2  = d\rho^{2} + \alpha^2 \rho^{2}d\theta^{2}, \label{metr1}
\end{equation}
where $\alpha=n_o/n_e$ is the ratio between the refractive indices. The geodesic equation in a Riemannian space like the cone is \cite{man}
\begin{equation}
\frac{d^{2}x^i}{dt^2}+\sum_{j,k}\Gamma^{i}_{jk}\frac{dx^j}{dt}\frac{dx^k}{dt}=0,\label{georie}
\end{equation}
where $t$ is a parameter along the geodesic and $\Gamma^{i}_{jk}$ are the Christoffel symbols, given by
\begin{equation}
\Gamma^{i}_{jk}=\frac{1}{2}\sum_{m}g^{mi}\left\{ \frac{\partial g_{km}}{\partial x^j}+\frac{\partial g_{mj}}{\partial x^k}-\frac{\partial g_{jk}}{\partial x^m}\right\} . \label{chris}
\end{equation}
For metric (\ref{metr1}) equation (\ref{georie}) reduces to the coupled system of ordinary differential eaquations
\begin{equation}
 \frac{d^2\rho}{dt^2}-\alpha^2\rho\left(\frac{d\theta}{dt}\right)^2=0 \label{eq1}
\end{equation}
and
\begin{equation}
 \frac{d^2\theta}{dt^2}+\frac{2}{\rho}\frac{d\rho}{dt}\frac{d\theta}{dt}=0. \label{eq2}
\end{equation}

The solution to the coupled system (\ref{eq1}) and (\ref{eq2}) is easily obtained \cite{padua}:
\begin{equation}
 \rho(t)=\sqrt{\frac{C^2}{E\alpha^2}+2E(t+D)^2}, \label{r(t)}
\end{equation}
\begin{equation}
 \theta(t)=\frac{1}{\alpha}\arctan\left(\frac{2E\alpha(t+D)}{c}\right)+\frac{F}{\alpha}, \label{theta(t)}
\end{equation}
where $C$, $D$, $E$ and $F$ are integration constants.

In figure 1 we show the light paths in the nematic medium with the $k=1$, $c=\frac{\pi}{2}$ disclination as given by (\ref{r(t)}) and (\ref{theta(t)}). In figure 2 the geodesics on a cone are shown for comparison.
\begin{figure}[!h]
\begin{center}
\includegraphics[height=5cm]{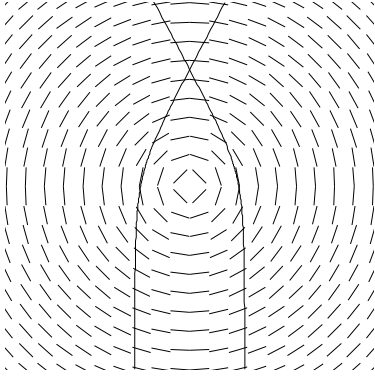} 
\caption{Light trajectories in a nematic liquid crystal with a topological defect given by a disclination  $k=1$ and $c=\pi/2$.}
\end{center}
\end{figure}
\begin{figure}[!h]
\begin{center}
\includegraphics[height=5cm]{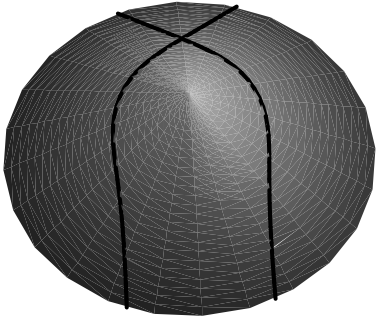} 
\caption{Geodesics on the cone.}
\end{center}
\end{figure}

Metric (\ref{metr1}) describes a cone. Fig. 3 shows the making of a cone from a planar sheet where an angular section was removed with posterior identification of the edges. If $\gamma$ is the angle that defines the removed section then the remaining surface corresponds to an angular sector of $2\pi\alpha=2\pi-\gamma$. This is exactly what metric (\ref{metr1}) describes. The incorporation of the term $\alpha^2$ to the Euclidean metric in polar coordinates makes the total angle on the surface be $\int_{0}^{2\pi} \alpha d\theta=2\pi\alpha <2\pi$, since $n_o<n_e$. It is clear then that $\alpha$ tells how ``pointed'' is the cone. The closer $\alpha$ gets to 1 the flatter is the cone. For $\alpha=1$ the cone turns into a plane.

\begin{figure}[!h]
\begin{center}
\includegraphics[height=1.5cm]{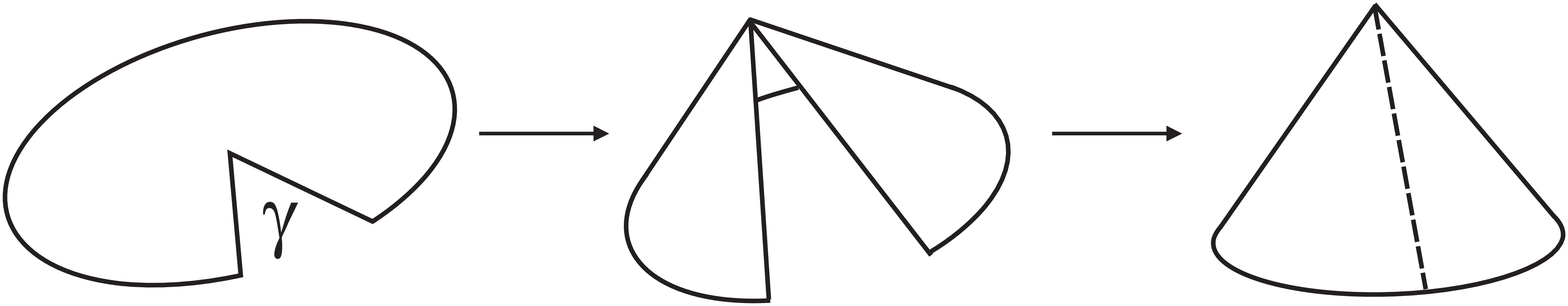} 
\caption{Conical surface of angular deficit $\gamma$.}
\end{center}
\end{figure}

The solutions (\ref{r(t)}) and (\ref{theta(t)}) still hold for the radial defect with $k=1$ and $c=0$ since, in this case, the line element (\ref{kmetric}) reduces to
\begin{equation}
ds^2  = d\rho^{2} + \frac{1}{\alpha^2} \rho^{2}d\theta^{2}, \label{metr2}
\end{equation}
where, $\alpha=n_o/n_e$ still. Consequently, equations (\ref{r(t)}) and (\ref{theta(t)}) become
\begin{equation}
 \rho(t)=\sqrt{\frac{C^2 \alpha^2}{E}+2E(t+D)^2}, \label{r(t)2}
\end{equation}
\begin{equation}
 \theta(t)=\alpha\arctan\left(\frac{2E(t+D)}{c\alpha}\right)+F\alpha, \label{theta(t)2}
\end{equation}
where, as before, $C$, $D$, $E$ and $F$ are integration constants.

\section{Refractive index variation}
The refractive indices $n_o$ and $n_e$ of a nematic liquid crystal depend both on the temperature ($T$) and on the wavelenth ($\lambda$) of the light. In this section, based in \cite{jun1} and \cite{jun2}, we analyse how these parameters affect the ratio $\alpha=n_o/n_e$, which characterizes the effective geometry associated to disclinations. By changing either $T$ or 
$\lambda$, $\alpha$ is changed and so is the effective geometry. This causes a deformation of the geodesics associated to the light rays in our model. 

In \cite{jun1} we can find expressions to the ordinary and extraordinary refractive index given in terms of the birefringence $\Delta n$ and of its average value $\left\langle n\right\rangle$, such that
\begin{equation}
n_o=\left\langle n\right\rangle-\frac{1}{3}\Delta n,\label{a1}
\end{equation}
\begin{equation}
n_e=\left\langle n\right\rangle+\frac{2}{3}\Delta n.\label{a2}
\end{equation}
In $(\ref{a1})$ and $(\ref{a2})$ the behavior of $\left\langle n\right\rangle$ as function of the temperature \cite{jun1} is given  through a linear dependence given by  
\begin{equation}
\left\langle n\right\rangle=A-BT,\label{med}
\end{equation}
where the parameters $A$ and $B$ are obtained experimentally.

The birefringence can be written in terms of the approximated \cite{haller} order parameter $S=\left(1-\frac{T}{T_c}\right)^{\beta}$ as 
\begin{equation}
\Delta n=(\Delta n)_0\left(1-\frac{T}{T_c}\right)^{\beta},\label{birre}
\end{equation}
where $(\Delta n)_0$ is the birefringence at $T=0\,K$, $\beta$ is a constant associated to the material and $T_c$ is the isotropic-nematic transition temperature. 

Therefore, substituting the equations $(\ref{med})$ and $(\ref{birre})$ into $(\ref{a1})$ and $(\ref{a2})$, we have
\begin{equation}
n_o=A-BT-\frac{(\Delta n)_0}{3}\left(1-\frac{T}{T_c}\right)^{\beta},\label{ind1}
\end{equation}
\begin{equation}
n_e=A-BT+\frac{2(\Delta n)_0}{3}\left(1-\frac{T}{T_c}\right)^{\beta}.\label{ind2}
\end{equation}

The liquid crystal considered was the 5CB or 4-cyano-4-n-pentylbiphenyl and the wavelength of the incident beam was $589$nm \cite{jun1}. For this material the parameters are given in the table below obtained from \cite{jun1}. The parameters A,  $\beta$ and $(\Delta n)_0$ are adimensional.

\begin{center}
\begin{tabular}{|c|c|c|c|c|} \hline
  A & B & $\beta$ &$(\Delta n)_0$ &$T_c$\\ \hline 
1.7546 & 0.0005360 K$^{-1}$ & 0.2391 & 0.3768 &306.6 K\\ \hline 
\end{tabular}
\end{center}

In figure 4 we show the ratio $\alpha=n_o/n_e$ as a function of the temperature as obtained from equations (\ref{ind1}) and (\ref{ind2}). As shown in \cite{caio2} these parameters are associated to the effective geometry perceived by the light traveling in the vicinity of $k=1$ disclinations. For these defects the geometry is conical with the radial disclination ($k=1$, $c=0$) behaving as a negative curvature cone and the vortex-like disclination ($k=1$, $c=\pi/2$) like the ordinary cone. The value $\alpha=1$, reached at $T_c$ corresponds to  the Euclidean geometry which describes the isotropic phase.

\begin{figure}[h]
\begin{center}
\includegraphics[height=7cm]{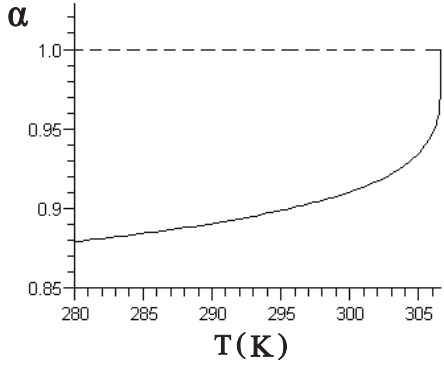} 
\caption{Effective geometry parameter $\alpha$  as function of the temperature for 5CB in the nematic phase at 589 nm.}
\end{center}
\end{figure}

Next, we consider the wavelength dependence of the effective geometry at a fixed temperature. Li and Wu \cite{jun2} modeled the ordinary and extraorinary refractive indices wavelength dependence based on the extended Cauchy formulae. Their model is described by the following equations:
\begin{equation}
n_e=A_e-\frac{B_e}{\lambda^2}+\frac{C_e}{\lambda^4},\label{lambda1}
\end{equation}
\begin{equation}
n_o=A_o-\frac{B_o}{\lambda^2}+\frac{C_o}{\lambda^4}.\label{lambda2}
\end{equation}

The coefficients appearing in equations (\ref{lambda1}) and (\ref{lambda2}) were obtained \cite{jun2} by fitting experimental data. For 5CB at 25.1 $^o$C they are given in the tables below.

\begin{center}
\begin{tabular}{|c|c|c|} \hline
  $A_e$ & $B_e$ & $C_e$ \\ \hline
1.6795 & 0.0048 $\mu m^2$& 0.0027 $\mu m^4$\\ \hline
\end{tabular}
\end{center}
\begin{center}
\begin{tabular}{|c|c|c|} \hline
  $A_o$ & $B_o$ & $C_o$\\ \hline
1.5174 & 0.0022 $\mu m^2$ & 0.0011 $\mu m^4$\\ \hline
\end{tabular}
\end{center}

In figure 5 we show the ratio $\alpha=n_o/n_e$ as a function of the wavelength as obtained from equations (\ref{lambda1}) and (\ref{lambda2}).

\begin{figure}[!h]
\begin{center}
\includegraphics[height=7cm]{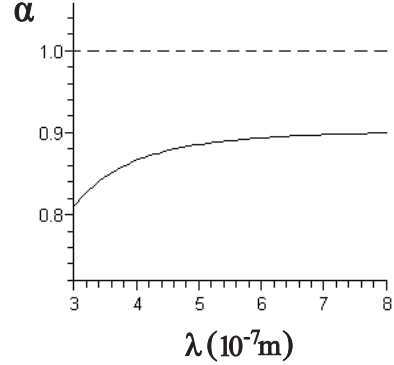} 
\caption{Effective geometry parameter $\alpha$  as function of the wavelength for 5CB  in the nematic phase at 25.1 $^o$C.}
\end{center}
\end{figure}

Since both temperature and wavelength cause $\alpha$ to change we can summarize their effect on the light paths by studying the geodesics for different values of $\alpha$. Substituting the metric (\ref{kmetric}) in (\ref{chris}) and this one in (\ref{georie})   we can calculate the geodesics for different values of $\alpha$. As described in Section II, the geodesic equations (\ref{georie}) have exact solutions for the $k=1$ case. The remaining cases  can be solved by a numerical method. In figures 6 and 7  we show the effects of the variation of the parameter $\alpha$ on the light paths near the $k=1$ defects, using the exact solution of Section II. In figure 8 we show the same effects for the $k=-1$ disclination, using the Runge-Kutta numerical method to solve the geodesic equation. In all cases, the solid line corresponds to $\alpha$=0,8912, the dotted line to $\alpha$=0,9120 and the dash-dotted line to $\alpha$=0,9355. These values, for 5CB, probed by a 589 nm light beam, correspond to the temperatures of 290 K, 300 K and 305 K, respectively. Notice that as $\alpha$ approaches 1 the light paths straighten out, as it should.

\begin{figure}[!h]
\begin{center}
\includegraphics[height=5cm]{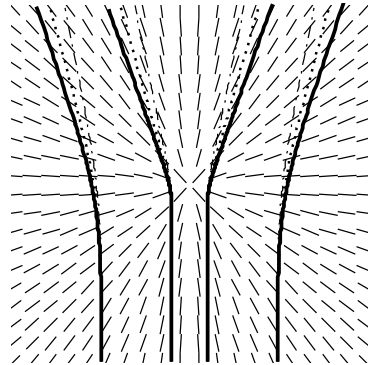} 
\caption{Influence of the parameter $\alpha$ on the light trajectories in a nematic liquid crystal with a disclination  $k=1$ and $c=0$.}
\end{center}
\end{figure}
\begin{figure}[!h]
\begin{center}
\includegraphics[height=5cm]{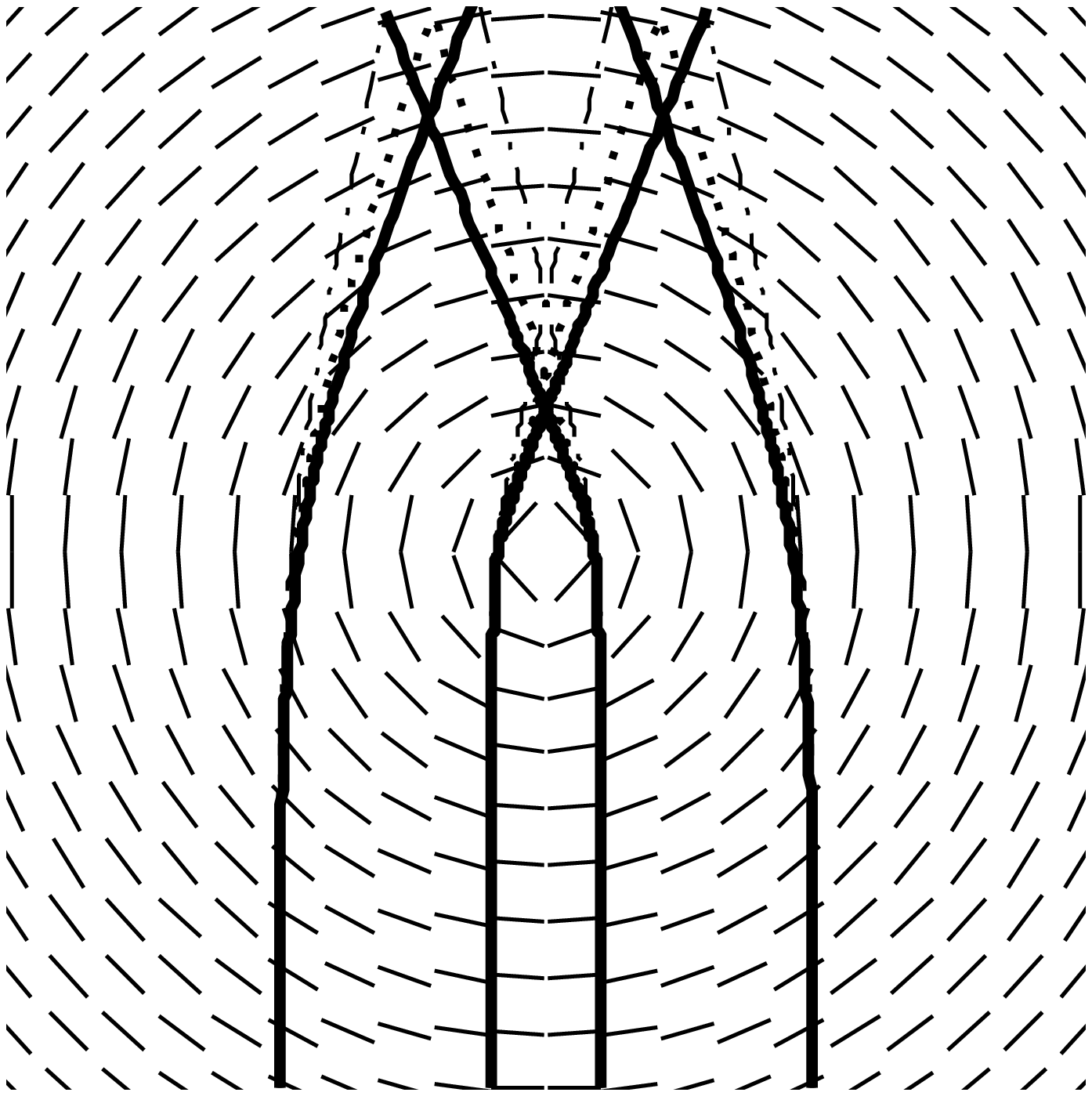} 
\caption{Influence of the parameter $\alpha$ on the light trajectories in a nematic liquid crystal with a disclination  $k=1$ and $c=\pi/2$.}
\end{center}
\end{figure}
\begin{figure}[!h]
\begin{center}
\includegraphics[height=5cm]{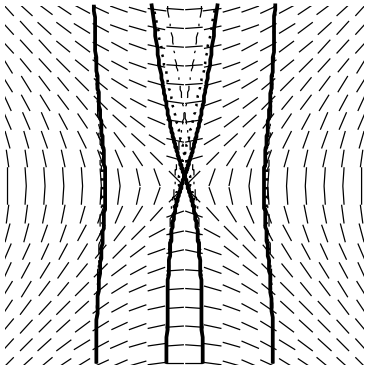} 
\caption{Influence of the parameter $\alpha$ in the light trajectories in a nematic liquid crystal with a disclination  $k=-1$ and $c=\pi/2$.}
\end{center}
\end{figure}

\newpage

\section{Conclusion}
Topological defects in nematics cause light passing by to deflect, as shown in \cite{caio1}. In that article, we associated the light paths to geodesics in a curved space specified by the defect. The deflection is due to the particular orientation of the director field associated to the defect, which may be translated into curvature. The intensity of the deflection depends on the ratio $\alpha$ between the ordinary and extraordinary refractive indices, which, in turn, depend on the temperature of the liquid crystal and on the wavelength of the light. Taking as example 5CB, which has been extensively characterized with respect to temperature and wavelength dependence  of the refractive indices \cite{jun1,jun2}, we solved the geodesic equations for a realistic range of values of $\alpha$ corresponding to temperature and/or wavelength variation. The graphical result illustrates the influence of these parameters on the light deflection caused by the defect. The further $\alpha$ gets from 1 the strong is the deflection. This can be achieved by either lowering the temperature or shortening the wavelength. In conclusion, the study of the influence of measurable physical parameters, like temperature and wavelength, helps us to understand better the behavior of light propagation in liquid crystals where topological defects are relevant.

\begin{acknowledgement}
This work has been supported by CNPq, CNPq/FACEPE, PRONEX/FAPESQ-PB and CAPES/PROCAD. We are indebted to Eduardo R. da Costa for helping with the graphs.
\end{acknowledgement}

\end{document}